\documentclass[epj]{svjour}
\usepackage{graphics}
\usepackage[pdftex]{graphicx}
\usepackage{epsfig}
\usepackage{epstopdf}
\usepackage[T1]{fontenc}
\usepackage{amsmath}

\usepackage[pdftex]{graphicx}
\usepackage[T1]{fontenc}

\usepackage{lineno}

\begin{document}

\title{New experimental limits on the $\alpha$ decays of lead isotopes \\}
\author{
J.~W.~Beeman\inst{1}, 
F.~Bellini\inst{2,3},
L.~Cardani\inst{2,3},
N.~Casali\inst{4,5},
S.~Di~Domizio\inst{6},
E.~Fiorini\inst{7,8},
L.~Gironi\inst{7,8},
S.S.~Nagorny\inst{5,9},
S.~Nisi\inst{5},
F.~Orio\inst{3},
L.~Pattavina\inst{8}\thanks{\emph{Corresponding Author} \newline \emph{e-mail address}: luca.pattavina@mib.infn.it},
G.~Pessina\inst{8},
G.~Piperno\inst{2,3},
S.~Pirro\inst{8},
E.~Previtali\inst{7,8},
C.~Rusconi\inst{8},
C.~Tomei\inst{3},
M.~Vignati\inst{3}
}

\institute{
Lawrence Berkeley National Laboratory, Berkeley, California 94720, USA  					\and 
Dipartimento di Fisica, Sapienza Universit\`{a} di Roma,  I 00185 Roma, Italy 				\and 
INFN, Sezione di Roma, I 00185 Roma, Italy 		   		  		  							\and 
Dipartimento di Fisica, Universit\`{a} degli studi dell'Aquila,  I 67100  L'Aquila, Italy 	\and 
INFN, Laboratori Nazionali del Gran Sasso, I 67010 L'Aquila, Italy 		  					\and 
INFN,  Sezione di Genova,   I 16146 Genova,  Italy 			 								\and 
Dipartimento di Fisica, Universit\`{a} di Milano-Bicocca, I 20126 Milano, Italy 			\and 
INFN, Sezione di Milano Bicocca, I 20126 Milano, Italy	  		  		  					\and 
Institute for Nuclear Research, MSP, 03680 Kyiv, Ukraine
}

%
%
\abstract{
For the first time a PbWO$_4$ crystal was grown using ancient Roman lead and it was run as a cryogenic detector. Thanks to the simultaneous and independent read-out of heat and scintillation light, the detector was able to discriminate $\beta$/$\gamma$ interactions with respect to $\alpha$ particles  down to low energies.
New more stringent limits on the $\alpha$ decays of the lead isotopes are presented.
In particular a limit of \mbox{T$_{1/2}$ $>$~1.4$\cdot$10$^{20}$~y} at a 90\% C.L. was evaluated for 
the $\alpha$ decay of $^{204}$Pb to $^{200}$Hg. 
\PACS{
      {23.60.+e}{$\alpha$ decay}   \and
      {29.40.Mc}{Scintillation detectors} \and 
	  {07.20.Mc}{Cryogenics; low-temperature detectors}
     } 
} 
\authorrunning{J.W.~Beeman \it{et al.}}
\maketitle
\section{Introduction}
\label{intro}
After the first observation of $^{209}$Bi~\cite{209Bi-nature} $\alpha$ decay, and the recent measurements of the half-life value including the transition to ground and to the first excited state~\cite{209Bi-nostro}, lead is considered to be the heaviest stable element. 
However $\alpha$ decay in lead is energetically allowed for all the four naturally occurring isotopes. 
The Q-values of the decays (Q$_\alpha$) and natural abundances ($\delta$) are listed in Tab.~\ref{Tab-1}, as well as calculations of half-lives based on the cluster models~\cite{cluster-1991,cluster-1992}, phenomenological fission theory of alpha decay~\cite{fission-1983}, semiclassical WKB approximation~\cite{WKB-1992}, and microscopic approach~\cite{micro-1988}. 
Given the values in Tab.~\ref{Tab-1}  there is no feasible perspective to observe the $\alpha$ decay of  $^{206}$Pb, $^{207}$Pb and $^{208}$Pb, but the measurements described in the present work might test the reliability of the various nuclear models.
The same perspective holds also for $^{204}$Pb , but in addition, this isotope plays also an important role in lead geochronology, as a standard reference~\cite{geo}.\newline

In 1958  there was a claim~\cite{Riezler-1958} of a possible $\alpha$ decay of  $^{204}$Pb. 
The nuclear emulsion technique with enriched (27.0\%) $^{204}$Pb was used for this experiment. A peak was found between 8~$\mu$m and 9~$\mu$m in the emulsions~\cite{Faraggi-1951}, corresponding to an $\alpha$ particle of $\approx$~2.6 MeV. This result 
was excluded by the mass data on $^{204}$Pb and $^{200}$Hg~\cite{toimass}, and the same measurement was then translated to the present quoted limit: T$_{1/2}>$ 1.4$\cdot$10$^{17}$~y.\newline

The need of increasing the experimental sensitivity to rare $\alpha$ decays lead to the employment of different and new experimental techniques such as ionizing chambers~\cite{ionizing-1961}, crystal scintillators~\cite{scint-2003} and liquid scintillators~\cite{liquid-2009}.\newline

Recently, measurements of $\alpha$ decays of  $^{209}$Bi~\cite{209Bi-nature,209Bi-nostro} and $^{180}$W~\cite{180W-Cresst}, with half-lives of 1.9$\cdot$10$^{19}$~y and 1.8$\cdot$10$^{18}$~y respectively, demonstrated the high sensitivity for the discovery of rare nuclear processes that can be achieved with scintillating bolometers. The main advantage of this technology is the wide choice of detector materials that allows to investigate isotopes that are not easily measurable with conventional detectors. Moreover, the simultaneous read-out of light and heat signals results in a powerful tool for background identification, thanks to the different amount  of light emitted by $\alpha$ and $\beta/\gamma$ particles of the same energy. This is fundamental in case the expected energy of an $\alpha$ decay lies in the environmental $\beta/\gamma$ energy region (dominated by the 2615~keV $\gamma$ line of $^{208}$Tl). 

\begin{table}
\label{Tab-1}
\caption{$\alpha$ transitions in Pb isotopes to the ground state of daughter nuclei. Q$_{\alpha}$  values are taken 
from~\cite{AUDI-2003}, natural isotopic abundances ($\delta$) from ~\cite{bohlke-2005}. The theoretical expected half-lives are evaluated using theoretical models given in~\cite{cluster-1991,cluster-1992,fission-1983,WKB-1992,micro-1988}.} 
\begin{center}
\begin{tabular}{lccc}
\hline\noalign{\smallskip}
Isotope &		  $\delta$ 		 & Q$_{\alpha}$    		  & T$^{theor}_{1/2}$ \\
		& 		  [\%] 		     & [keV] 				  &		[y]			  \\
\noalign{\smallskip}\hline\noalign{\smallskip}
$^{204}$Pb  & 1.4~(1)  & 1969.5~(12)	 & 2.3$\cdot$10$^{35}  \div$ 1.2$\cdot$10$^{37}$ \\
$^{206}$Pb  & 24.1~(1) & 1135.5~(11) & 1.8$\cdot$10$^{65}  \div$ 6.7$\cdot$10$^{68}$  \\
$^{207}$Pb  & 22.1~(1) & 392.3~(13) & 3.6$\cdot$10$^{152} \div$ 3.4$\cdot$10$^{189}$  \\
$^{208}$Pb  & 52.4~(1) & 516.9~(13) & 1.2$\cdot$10$^{124} \div$ 7.4$\cdot$10$ ^{132}$  \\
\noalign{\smallskip}\hline
\end{tabular}
\end{center}
\end{table}

\section{PbWO$_4$ crystal from ancient Roman lead}
\label{roman-crystal}
When coming to low background measurements, the main problem of lead is the high activity in $^{210}$Pb. The radioactivity of ``commercial'' lead can be of the order of few tens of Bq/kg,  for ore-selected samples~\cite{lead210-activity-1}, while on other samples it can reach an activity of thousands of Bq/kg~\cite{lead210-activity-2}.\newline

Besides the excellent performances that characterize thermal detectors, a limiting factor arises from the relatively slow time response of these devices. 
Depending on the technology of the temperature sensors, as well as on the dimension of the absorber, the time development of the thermal pulse can range between few $\mu$s up to few seconds. This last point has to be taken into account in the case of crystal compounds based on lead.\newline
$^{210}$Pb is a decay product of the $^{238}$U decay chain, which is present in all rocks and ores. During the melting of the ore, $^{210}$Pb concentrates  in the lead-metal, while all the other long-lived radioactive nuclides of the $^{238}$U chain ( $^{234}$U, $^{230}$Th and $^{226}$Ra) are extracted from the slag since they are chemically different from Pb. Since the $^{210}$Pb half-life is 22.3 y, its activity should be extremely small in ancient samples. For this reason ancient Roman lead~\cite{Piombo-Romano} was used to grow our crystal. This lead, in form of ingots, was recovered from the wreck of a Roman ship sunk near the coast of Sardinia in the Mediterranean Sea.
The $^{210}$Pb content of this Roman lead was measured to be less than 4~mBq/kg. 
We decided to grow a standard lead containing scintillator like a  PbWO$_4$ crystal~\cite{PbWO4}.
A small sample of this lead was sent to the Preciosa company in Hungary, to grow an undoped PbWO$_4$ crystal. 
The request to have a clean and undoped crystal arises from two different considerations:
\begin{itemize}
\item in a bolometer, the thermal signal strongly depends upon the crystal purity; 
\item the scintillation yield of doped crystals normally decreases at very low temperatures, while for pure crystals it increases~\cite{Birks} .
\end{itemize}

\begin{table}
\label{Tab-2}
\caption{Lead isotopic abundaces for the PbWO$_4$ crystal, evaluated with ICP-MS measurements. The values are in good agreement with the natural abundances quoted in Tab.~\ref{Tab-1}.} 
\begin{center}
\begin{tabular}{cccc}
\hline\noalign{\smallskip}
$^{204}$Pb	   &	$^{206}$Pb	  &   $^{207}$Pb       &    $^{208}$Pb       \\ \noalign{\smallskip}\hline\noalign{\smallskip}
1.34$\pm$0.02~\%  &  25.10$\pm$0.25~\%  & 21.1$\pm$0.30~\%   &     52.4$\pm$0.50~\%     \\
\noalign{\smallskip}\hline
\end{tabular}
\end{center}
\end{table}

The PbWO$_4$ crystal used in this work has dimensions of 3.0$\times$3.0$\times$6.1~cm$^3$ and a total mass of 454.1~g.
A tiny splint ($\sim$ 50~mg) was removed from the crystal and analyzed through ICP Mass Spectroscopy in order to evaluate the isotopic abundances of the four lead isotopes. The results are shown in Tab.~\ref{Tab-2}.

\section{Experimental Set-up}
\label{Set-Up}

\begin{figure}
\centering
\resizebox{0.4\textwidth}{!}{%
  \includegraphics{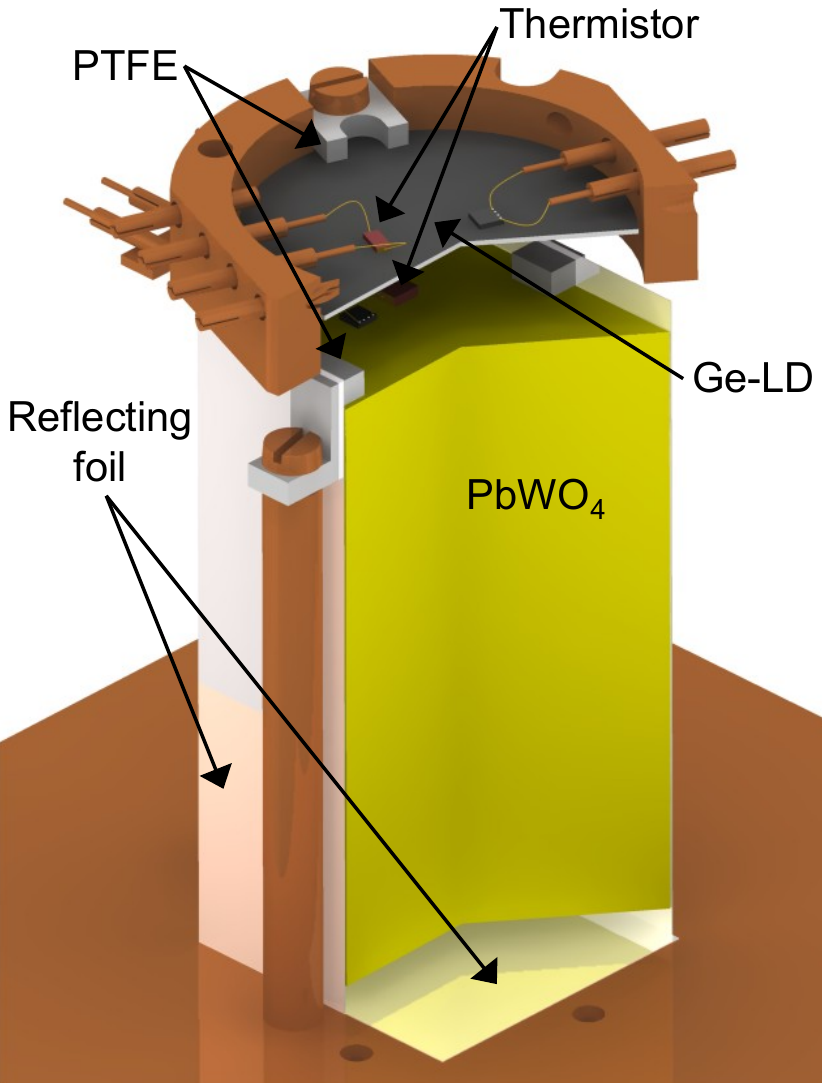}}
\caption{Set-up of the detector. Two Cu columns, one fixing  the S-shaped PTFE and one fixing the LD Cu frame are not visible due to 
the chosen cross section.}
\label{fig:1}   
\end{figure}

The scheme of the set-up is shown in Fig.~\ref{fig:1}.
The PbWO$_4$ crystal is held by means of four S-shaped PTFE supports fixed to four cylindrical Cu columns. It is  surrounded (with no direct contact) by a plastic reflecting foil (3M  VM2002). The Light Detector (LD)~\cite{NIMA-2006-A}  is constituted by a  36~mm diameter, 1~mm thick pure Ge crystal absorber working as a bolometer: it is heated up by the  absorbed photons  and the temperature variation is proportional to the scintillation signal.  
The Ge wafer is held by two custom-shaped PTFE holders embedded in a  cylindric Cu frame. 
The frame is held above the PbWO$_4$ crystal using two Cu columns.  

The temperature sensor of the  PbWO$_4$ crystal is a 3$\times$3$\times$1~mm$^3$ Neutron Transmutation Doped (NTD) germanium thermistor, the same used 
in the Cuoricino experiment~\cite{QINO}. It is thermally coupled to the crystal via 9 glue spots of $\approx$~0.6~mm diameter and $\approx$~50~$\mu$m height.
The temperature sensor of the LD has a smaller volume (one third of the PbWO$_4$'s one) in order to decrease its heat capacity, increasing therefore its 
thermal signal.
The PbWO$_4$ detector was run  in an Oxford 200 $^3$He/$^4$He dilution cryostat deep underground in the Laboratori Nazionali del Gran Sasso.
The details of the electronics and of the cryogenic facility can be found elsewhere \cite{NIMA-2006-B,NIMA-2006-C}.

The heat and light pulses, produced by interacting particles in the absorber (transduced in a voltage pulse by the NTD thermistors) are amplified and fed into an 18-bit NI-6284 PXI ADC unit.

The trigger is software generated on each bolometer and when it fires waveforms 1~s long, sampled at 1 kHz,  are saved on disk. 
Moreover, when the trigger of the  PbWO$_4$ crystal fires, the corresponding waveform from the LD is recorded,  irrespective of its trigger.

The amplitude and the shape of the voltage pulse is then determined by the off-line analysis.

The heat axis is energy-calibrated attributing to each identified peak (due to an external $\gamma$ source and to internal $\alpha$ contaminations) the nominal energy of the line. Two independent functions were used for calibrating the $\alpha$ and $\beta$/$\gamma$ bands.

The dependency of amplitude from energy is parameterized with a second order polynomial of the heat pulse amplitude.

The energy calibration of the LD is obtained by means of a weak $^{55}$Fe source placed close to the Ge wafer that illuminates homogeneously the face opposed to the  PbWO$_4$ crystal.

\section{Data Analysis}
\label{Data-analysis}
\begin{figure}
\centering
\resizebox{0.5\textwidth}{!}{%
  \includegraphics[angle=0]{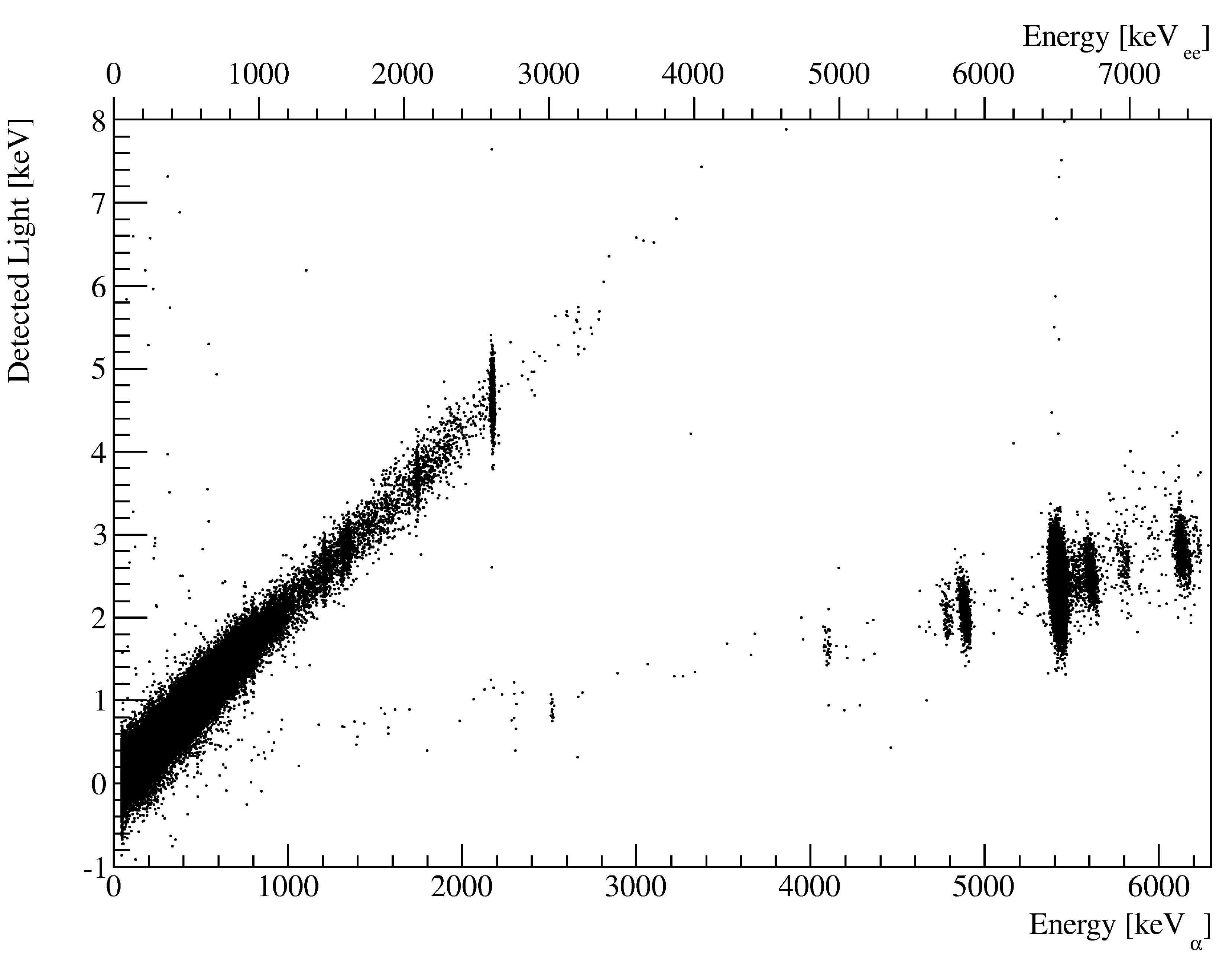}}
\caption{Light vs Heat scatter plot corresponding to 586 h of measurement. The horizontal axis reported on the bottom of the plot corresponds to an $\alpha$ energy calibration, the one on the top a $\beta$/$\gamma$ calibration.} 
\label{fig:2}   
\end{figure}

Our detector was operated for a total live time of 586 hours for a background measurement. In Fig.~\ref{fig:2} we present the obtained Light vs Heat scatter plot, with the two different energy calibrations.

To maximize the signal-to-noise ratio, the pulse amplitude is estimated by means of the Optimum Filter technique (OF)~\cite{GattiManfredi,Radeka:1966}. The filter transfer function is built from the ideal signal shape $s(t)$ and the noise power spectrum $N(\omega)$. The $s(t)$ is estimated by averaging a large number of triggered pulses (so that stochastic noise superimposed to each pulse averages to zero) while $N(\omega)$ is computed averaging the power spectra of randomly acquired waveforms where no pulse was found. The amplitude of a signal is estimated as the maximum of the filtered pulse. The amplitude of the scintillation light signals, instead, is evaluated from the filtered waveforms at a fixed time delay with respect to the PbWO$_4$ bolometer, as described in detail in~\cite{light_sync}.\newline

After the application of the OF, signal amplitudes are corrected for temperature and gain instabilities of the set-up. The PbWO$_4$ bolometer is calibrated with the most intense $\alpha$ peaks from the internal $^{238}$U and $^{232}$Th contaminations, and with an external $^{232}$Th and $^{40}$K $\gamma$ source, see Fig.~\ref{fig:2}. Since we are investigating low energy $\alpha$ decays, the internal $\alpha$ contaminations do extend over a large energy range, but they are not low enough to reach our region of interest (see Tab.~\ref{Tab-1}) in order to allow a thorough energy calibration.\newline

One of the constituent of the crystal is tungsten, the natural $\alpha$ decay of $^{180}$W at 2516~keV~\cite{180W-Cresst} is taken into account for the energy calibration.
On the other side, the LD is calibrated using the $^{55}$Fe X-ray doublet: 5.9~keV and 6.2~keV.\newline

The final spectrum is composed of events which survived two different types of data selection global and event-based requirements.
Global requirements are applied following criteria decided a priori on the detector performances (\emph{e.g.} excessive noise level). They identify bad time intervals that need to be discarded. Event-based requirements include: pile-up rejection and pulse-shape selection. The presence of a pile-up prevents the OF algorithm from providing a correct evaluation of the pulse amplitude. The pulse-shape analysis is used to reject non-physical events (\emph{e.g.} electronic spikes). The pulse-shape parameters used for selecting the events are the rise time and decay time of the OF-filtered waveform and the mean quadratic deviation of raw signals from the average detector response.

\subsection{Efficiency of event-based cuts}
Even if the crystal has been grown using ancient lead, the main background source is still $^{210}$Pb due to an internal crystal recontamination, as it will be explained in the Sect.~\ref{int_contaminations}. Because of the intense $^{210}$Po contamination, there is a significant loss of efficiency due to pile-up event rejection ($\epsilon_{pile-up}$). The efficiency is estimated as in~\cite{CCVR}:
\begin{equation}
\epsilon_{pile-up}\, = \, 1\, - \, P_{pile-up}\, =\, e^{-r \cdot T}
\end{equation}
where $P_{pile-up}$ is the pile-up probability, $r$ is the rate of the $^{210}$Po events that passed the global cuts ($r=$~66~mHz), $T$ the time window containing an event during which the occurrence of another one would be considered pile-up. The acquired waveform are 1~s long, so $T$ is set at 1~s.\newline
The pulse shape cut efficiency ($\epsilon_{PS}$) is estimated at different energies for various $\alpha$ lines, by a simultaneous fit on both spectra of accepted and rejected events~\cite{QINO}.
\begin{table}
\begin{center}
\caption{Efficiencies for $\alpha$ events that passed events-based cuts: pile-up, pulse-shape and total.} 
\begin{tabular}{lcccc}
\hline\noalign{\smallskip}
Nuclide & Q-value & $\epsilon_{pile-up}$ &  $\epsilon_{PS}$ & $\epsilon_{TOT}$ \\ 
\noalign{\smallskip}\hline\noalign{\smallskip}
$^{180}$W & 2516~keV & 0.94$\pm$0.01 & 0.88$\pm$0.02 & 0.83$\pm$0.02 \\
$^{232}$Th & 4082~keV & 0.94$\pm$0.01  & 0.87$\pm$0.01 & 0.81$\pm$0.02 \\
$^{230}$Th & 4770~keV & 0.94$\pm$0.01  & 0.85$\pm$0.01 & 0.80$\pm$0.02 \\
$^{222}$Rn & 5590~keV & 0.94$\pm$0.01  & 0.87$\pm$0.01 & 0.80$\pm$0.02 \\
$^{218}$Po & 6115~keV & 0.94$\pm$0.01  & 0.86$\pm$0.01 & 0.80$\pm$0.02 \\
$^{210}$Po & 5407~keV & 0.94$\pm$0.01  & 0.85$\pm$0.01 & 0.79$\pm$0.02 \\
\noalign{\smallskip}\hline
\end{tabular}
\label{Tab-3}

\end{center}

\end{table}
In Tab.~\ref{Tab-3} we report the values for the efficiencies of event-based cuts.

\subsection{Detector performances}

\begin{table}
\caption{Technical details for the PbWO$_4$ crystal and for the LD. Signal represents the absolute voltage drop across the thermistor for a unitary energy deposition, $\tau_R$ and $\tau_D$ are the rise and decay time, respectively.} 
\begin{center}
\begin{tabular}{lcccc}
\hline\noalign{\smallskip}
Crystal & Signal  & FWHM$_{base}$ &  $\tau_R$ & $\tau_D$ \\ 
            &  [$\mu$V/MeV]& [keV] &  [ms] & [ms]\\
\noalign{\smallskip}\hline\noalign{\smallskip}
PbWO$_4$ & 153 & 1.51 & 13 & 35    \\
LD & 1553 & 0.28 & 3 & 20\\
\noalign{\smallskip}\hline
\end{tabular}
\end{center}
\label{Tab-4} 
\end{table}

The detector performances are reported in Tab.~\ref{Tab-4}. The baseline resolution, FWHM$_{base}$, is governed by the noise fluctuations at the filtered output, and does not depend on the absolute pulse amplitude. The rise ($\tau_R$) and decay times ($\tau_D$) of the pulses are computed as the time difference between the 10$\%$ and the 90$\%$ of the leading edge, and the time difference between the 90$\%$ and 30$\%$ of the trailing edge, respectively.\newline

The FWHM energy resolution of the crystal ranges from 15.2$\pm$0.5~keV at 583~keV to 15.7$\pm$0.3~keV at 2615~keV for the $\gamma$ de-excitation of $^{208}$Tl. While for the $\alpha$ region it goes from 13.8$\pm$5.4~keV at 2516~keV ($^{180}$W) to 52.9$\pm$1.9~keV  at 6115~keV ($^{218}$Po), both lines refer to $\alpha$~+~recoil energies of the nucleus.\newline

The computed FHWM for the $^{55}$Fe on the light detector is 379$\pm$8~eV.

\subsection{Light Yields for $\alpha$ and $\beta$/$\gamma$ particles}
As described in Sect.~\ref{intro}, particle discrimination is the key point for all the experiments searching for rare decays in overwhelming background.\newline

In Fig.~\ref{fig:2}  we can distinguish the $\beta$/$\gamma$ and $\alpha$ regions which give rise to very clear separate distributions. In the upper band, ascribed to $\beta$/$\gamma$ events, the 2615~keV $^{208}$Tl $\gamma$-line is clearly visible. The lower band, populated by $\alpha$ decays, shows the continuum of events, all the way down to few keV, induced by the degraded $\alpha$ of $^{210}$Po, as well as $^{238}$U and $^{232}$Th contaminations.\newline

The Light Yield ($LY$), defined as the ratio between the measured light (in keV) and the nominal energy of an event (in MeV), was measured for the various $\gamma$-lines and it was found to have a constant value:
\begin{equation}\label{ly_beta}
LY_{\beta / \gamma} = 1.78 \pm 0.01~\mbox{keV/MeV}
\end{equation}
On the contrary, the $LY_{\alpha}$ for $\alpha$ decays has a value which is correlated to the nominal energy of the interacting particle.  Fitting the $\alpha$ band with a second order polynomial, we can estimate the $LY_\alpha$ for the various lines and we obtain the following relation:
\begin{equation}\label{ly_alpha}
LY_{\alpha}(E_\alpha)  = (0.28 \pm 0.01) + (2.93 \pm 0.14) \cdot 10^{-5} \cdot E_{\alpha} ~\mbox{keV/MeV}
\end{equation}
The observed behaviour reflects the fact that an $\alpha$ particle produces less scintillation light compared to an electron because of its larger $dE/dx$, which can induce saturation effects in the scintillator (the "Birks law"~\cite{Birks}). For the $\alpha$ band, the energy dependence of the $LY$ is a consequence of the energy dependence of the stopping power. Electrons, due to their low stopping power, do not suffer of saturation effects and their $LY$ is, consequently, energy independent.\newline

The scintillation Quenching Factor ($QF$) is defined as the ratio of the scintillation yield of an interacting particle ($\alpha$, neutron, nucleus) with respect to the $LY$ of a $\beta$/$\gamma$ event at the same energy. Since the $LY$ of $\alpha$ events varies with the energy of the interacting particle then the $QF$ for these type of events does not have a constant value. Using the eqs. \ref{ly_beta} and \ref{ly_alpha} we obtain a value of 0.20$\pm$0.01 for the $QF_\alpha$($^{180}$W). This value is in agreement with \cite{Tretyak}.\newline

In our measurement, we observe that two different functions are needed for the calibration of the two bands~\ref{cdwo4}, as it is shown in Fig.~\ref{fig:2}. In fact, if we apply the $\alpha$ calibration to the $\beta$/$\gamma$ band, we obtain a large mis-calibration, about 80\% for the $^{208}$Tl $\gamma$ line at 2615~keV. The explanation for this behavior relies in the mechanism of energy conversion~\cite{lecoq}. In fact, the energy deposited in the abosrber can be converted into:
\begin{itemize}
\item scintillation light, due to the excitation of luminescence centres;
\item thermal phonons;
\item "blind" channels, not detectable.
\end{itemize} 
Unfortunately, we are not able to measure the energy partition in the different channels, since we miss information on the light collection efficiency of the LD, and obviously we are not sensitive to the "blind" channel.

\subsection{Internal contaminations}
\label{int_contaminations}
Even if the starting materials for the crystal growth had a very high radiopurity level (Sect.~\ref{roman-crystal}) during the crystallization process is possible to unintentionally introduce some radioactive contaminants. These have been evaluated by analyzing the $\alpha$ region of the spectrum, due to the more favorable signal-to-noise ratio.\newline

\begin{figure}
\centering
\resizebox{0.5\textwidth}{!}{%
  \includegraphics{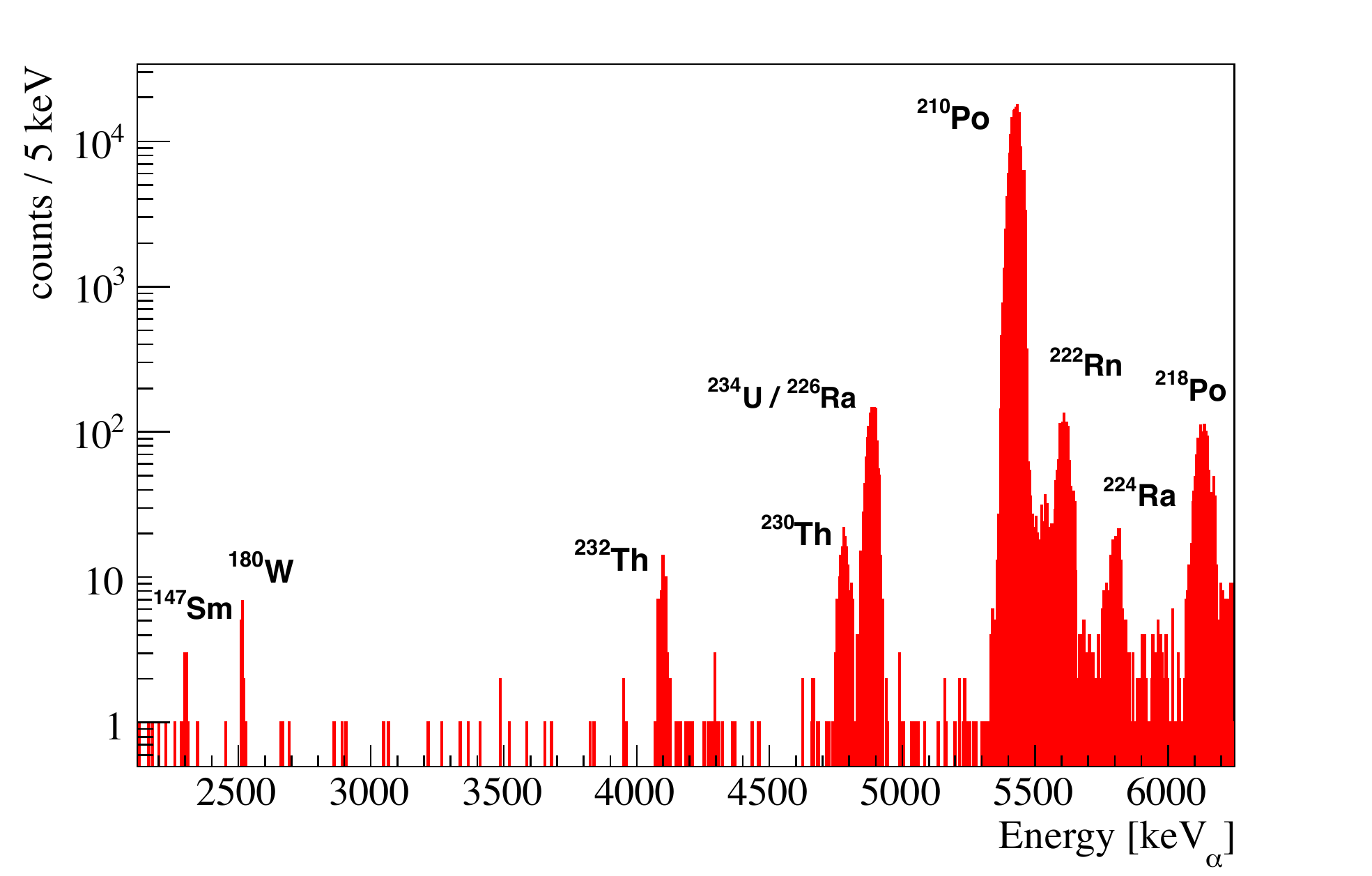}}
\caption{Energy spectrum of the $\alpha$ region, acquired over 586 h of measurement.} 
\label{fig:3}   
\end{figure}

As already pointed out, the background in this region is essentially due to $\alpha$ particles, which produce peaks and a continuum of events generated by degraded $\alpha$'s from the surfaces of the crystal and the surrounding materials~\cite{sticking}. This is mainly due to an intense $^{210}$Po contamination of the crystal as shown in Fig.~\ref{fig:3}.\newline

By means of Monte Carlo simulations, we estimate the background induced by a uniform $^{210}$Po contamination throughout the crystal. We obtain a flat background from zero energy up to the $^{210}$Po Q-value, the counting rate in this region is 0.02~counts/keV.

We observe that the PbWO$_4$ crystal is contaminated in $^{230}$Th and $^{210}$Po, from the $^{238}$U decay chain, $^{232}$Th (the whole $^{232}$Th decay chain) and $^{147}$Sm, see Tab.~\ref{Tab-5}.
\begin{table}
\caption{Evaluated internal radioactive contaminations for the PbWO$_4$ crystal. Limits are at 90\% C.L.} 
\begin{center}
\begin{tabular}{lcc}
\hline\noalign{\smallskip}
Chain & Nuclide  & Activity \\ 
            & & [$\mu$Bq/kg] \\
\noalign{\smallskip}\hline\noalign{\smallskip}
$^{232}$Th & $^{232}$Th & 51$\pm$8 \\
\noalign{\smallskip}\hline\noalign{\smallskip}
$^{238}$U & $^{238}$U & $<$~10 \\
 & $^{234}$U & $<$~7 \\
 & $^{230}$Th & 178$\pm$15 \\
 & $^{226}$Ra & 1403$\pm$43 \\
 & $^{210}$Po & (186$\pm$1)$\cdot$10$^3$ \\
\noalign{\smallskip}\hline\noalign{\smallskip}
$^{147}$Sm & $^{147}$Sm & 7$\pm$5 \\
\noalign{\smallskip}\hline
\end{tabular}
\label{Tab-5} 

\end{center}
\end{table}
For the $^{232}$Th decay chain we report just the activity of the progenitor of the chain, because the chain is at secular equilibrium: in fact the crystal was grown more than 14 years before the measurement, thus all the nuclides of the chain have the same activity. On the contrary for the $^{238}$U chain we evaluate the activity of all isotopes since the equilibrium is broken.

\subsection{Half-lives of lead isotopes} 

\begin{figure}
\centering
\resizebox{0.5\textwidth}{!}{%
  \includegraphics[angle=0]{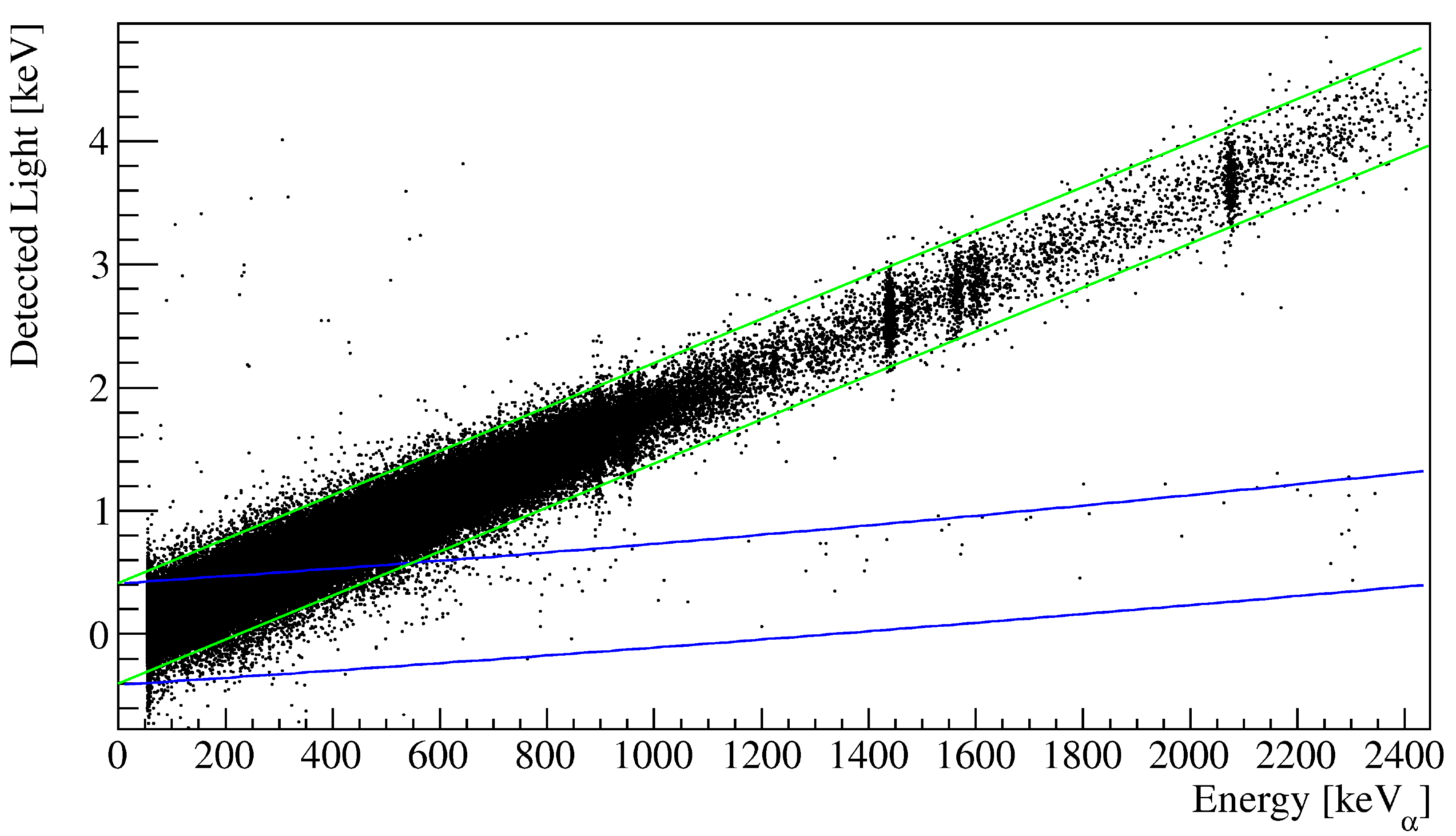}}
\caption{Low energy scatter plot. The green and blue lines represent the 3~$\sigma$ acceptance region, respectively, for $\beta$/$\gamma$ and $\alpha$ events.} 
\label{fig:bands}   
\end{figure}

We report in Tab.~\ref{Tab-1} all the lead isotopes that we have investigated with our PbWO$_4$ crystal. The $\alpha$/$\beta$ discrimination allows us to easily disentagle the $\alpha$ and $\beta$/$\gamma$ particle interactions. Unfortunately, the more we go at low energy the less efficient is the discrimination power due to the poor energy resolution of the LD. For energy releases in the PbWO$_4$ crystal smaller than 1~MeV, $\beta$/$\gamma$ events leak in the $\alpha$ region, because the two bands start to merge. Thus $\beta$/$\gamma$ events induce a negligible background for the investigation of $^{204}$Pb and $^{206}$Pb decays, but not for $^{207}$Pb and $^{208}$Pb, which are at lower energy. This background is therefore evaluated by defining the 3~$\sigma$ acceptance region for $\alpha$ and $\beta$/$\gamma$ events, as shown in Fig.~\ref{fig:bands}.

\subsubsection{$^{204}$Pb}
\label{Pb204}

\begin{figure}
\centering
\resizebox{0.5\textwidth}{!}{%
  \includegraphics{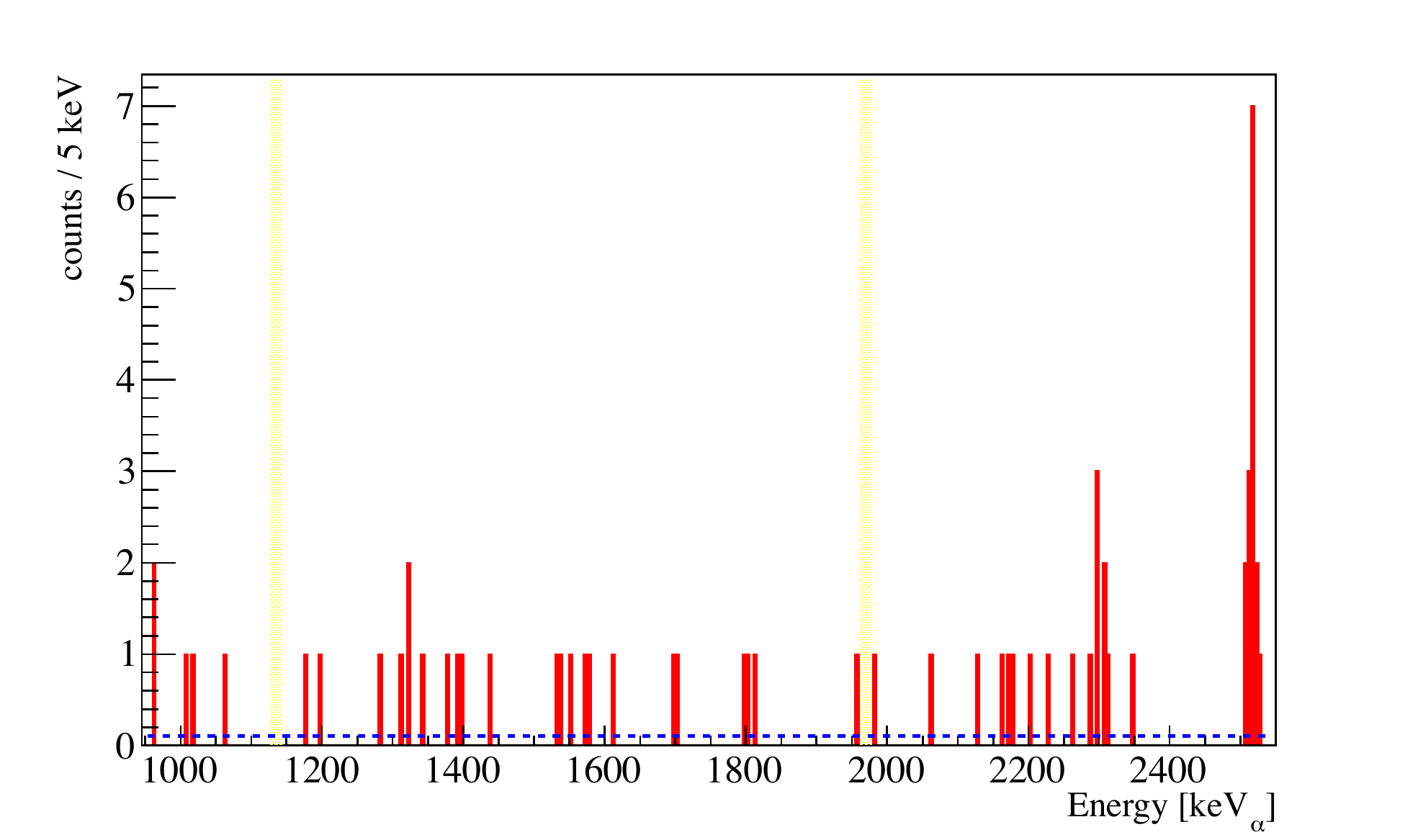}}
\caption{Energy spectrum of just $\alpha$ events, the dashed line represents the expected $^{210}$Po flat background. The ROI for $^{204}$Pb and $^{206}$Pb decays are highlighted.} 
\label{fig:5}   
\end{figure}

It is theorized that $^{204}$Pb can $\alpha$ decay on $^{200}$Hg, the Q-value of the transition is 1969.5~keV. The alpha energy spectrum in this energy region, shown in Fig.~\ref{fig:5} clearly shows no peak, but we have a flat background, as already mentioned in Sect.~\ref{int_contaminations}.  We study an interval of 2.8~$\sigma$ (that corresponds to a confidence interval of 99.5\%) centered around the Q-value of the transition, being $\sigma$ the energy resolution of the $^{180}$W peak (5.9~keV). The computed $^{210}$Po background around the 2.8~$\sigma$ region of interest (ROI) of 16.4~keV, is 0.3~counts, while the observed number of counts is 0. Applying the Feldman-Cousins method~\cite{Feldman} we are able to set a 90\% C.L. limit on the half-life of $^{204}$Pb, using the following formula:
\begin{equation}
T_{1/2}\, = \, N_{nuclei} \, \cdot \, \left( \frac{N_{counts}}{T_{meas}\, \cdot\, \epsilon_{TOT}} \right)^{-1}  \cdot \, \ln(2)
\end{equation} 
where: $N_{nuclei}$ is the number of nuclei of the analyzed isotope, $N_{counts}$ is the number of events estimated using the Feldman-Cousin approach, $T_{meas}$ is the measurement time in year and $\epsilon_{TOT}$ is the total efficiency of event-based cuts, which is assumed to have the same value of the $^{180}$W one. The computed lower-limit is:
\begin{equation*}
T^\alpha_{1/2}(^{204}Pb)\, >~1.4\cdot10^{20}~y \; \; \; \mbox{at \;90\% C.L.}
\end{equation*} 

\subsubsection{$^{206}$Pb}
\label{Pb206}

$^{206}$Pb is energetically allowed to decay through the $\alpha$ channel on $^{202}$Hg. The energy released in the decay is 1135.5~keV. Looking at Fig.~\ref{fig:5}, we can estimate the $^{206}$Pb half-life applying the same procedure explained in the previous section. The events distribution in the energy spectrum is still flat in the 1~MeV region, so no excess is observed at 1135.5~keV. The evaluated background is 0.2~counts, and the observed number of events in the 2.8~$\sigma$ energy interval (16.4~keV) around the Q-value is 0~counts. The lower-limit on the $^{206}$Pb half-life is:
\begin{equation*}
T^\alpha_{1/2}(^{206}Pb)\, >~2.5\cdot 10^{21}~y \; \; \; \mbox{at \;90\% C.L.}
\end{equation*}

\subsubsection{$^{207}$Pb}
\label{Pb207}
\begin{figure}
\centering
\resizebox{0.5\textwidth}{!}{%
  \includegraphics[angle=0]{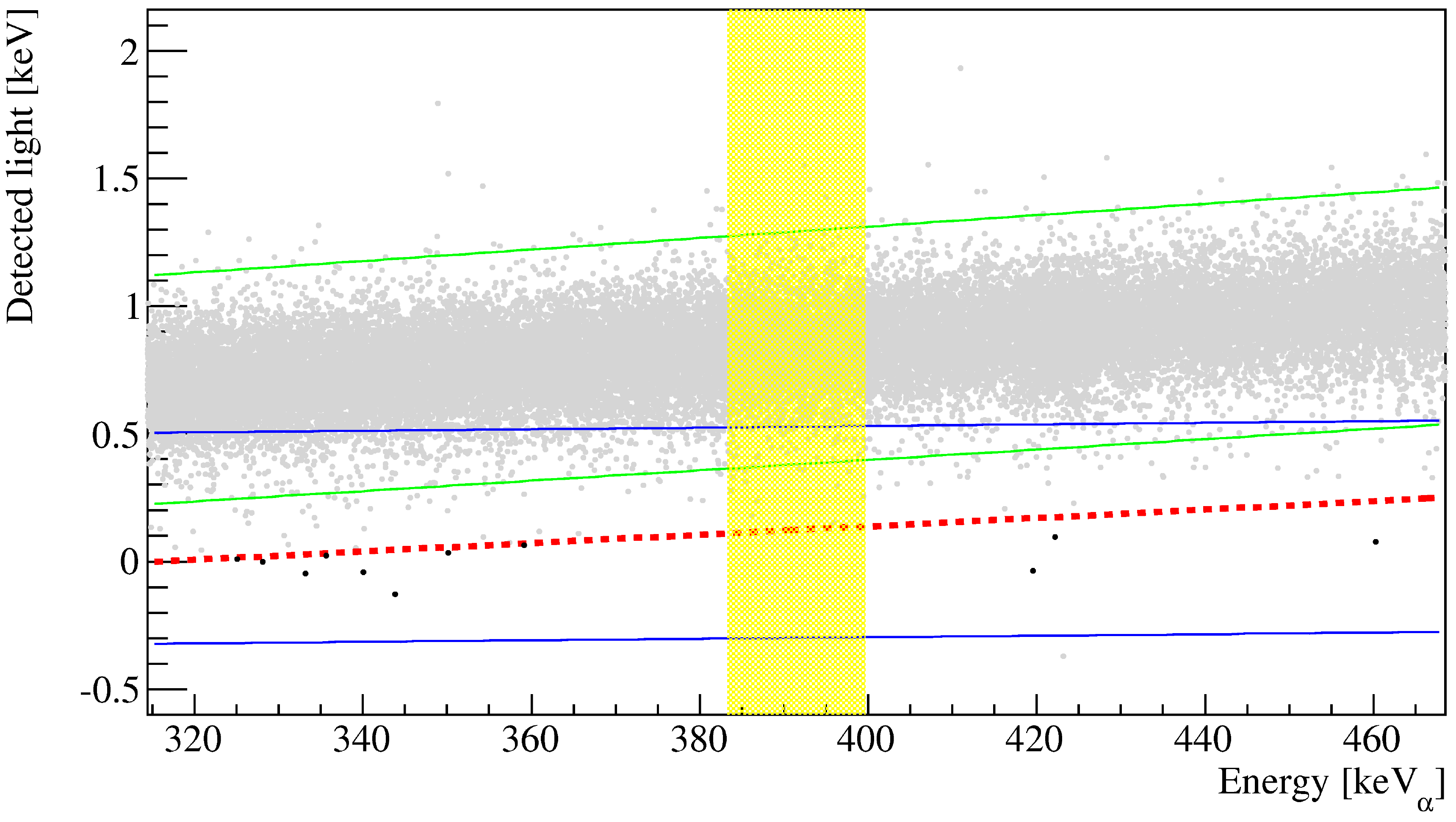}}
\caption{Low energy scatter plot analyzed for investigating $^{207}$Pb $\alpha$ decay. The blue and green lines represent the 90\%  acceptance region, respectively, for $\alpha$ and $\beta$/$\gamma$ events. The red dashed line is the energy cut applied for selecting the region with the highest statistical significance, while the shadowed region is our ROI.} 
\label{fig:cut300}   
\end{figure}

As already explained, for low energy $\alpha$ decays, where events from the $\beta$ band can give a contribution to the $\alpha$ region, we should define an energy cut that maximizes the signal-to-noise ratio at such low energy. In Fig.~\ref{fig:cut300} is shown the scatter plot around the Q-value: 392.3~keV, and the applied cut on the detected light that maximizes the statistical significance in that region. The estimated cut efficiency is 53\%.\newline
In the analyzed 2.8~$\sigma$ interval (16.4~keV) around the Q-value the number of observed events is 0 and the estimated background is 1~counts, therefore we can set a lower-limit on the $^{207}$Pb half-life:
\begin{equation*}
T^\alpha_{1/2}(^{207}Pb)\, >~1.9\cdot10^{21}~y \; \; \;\mbox{at \;90\% C.L.}
\end{equation*} 

\subsubsection{$^{208}$Pb}
\label{Pb208}
\begin{figure}
\centering
\resizebox{0.5\textwidth}{!}{%
  \includegraphics[angle=0]{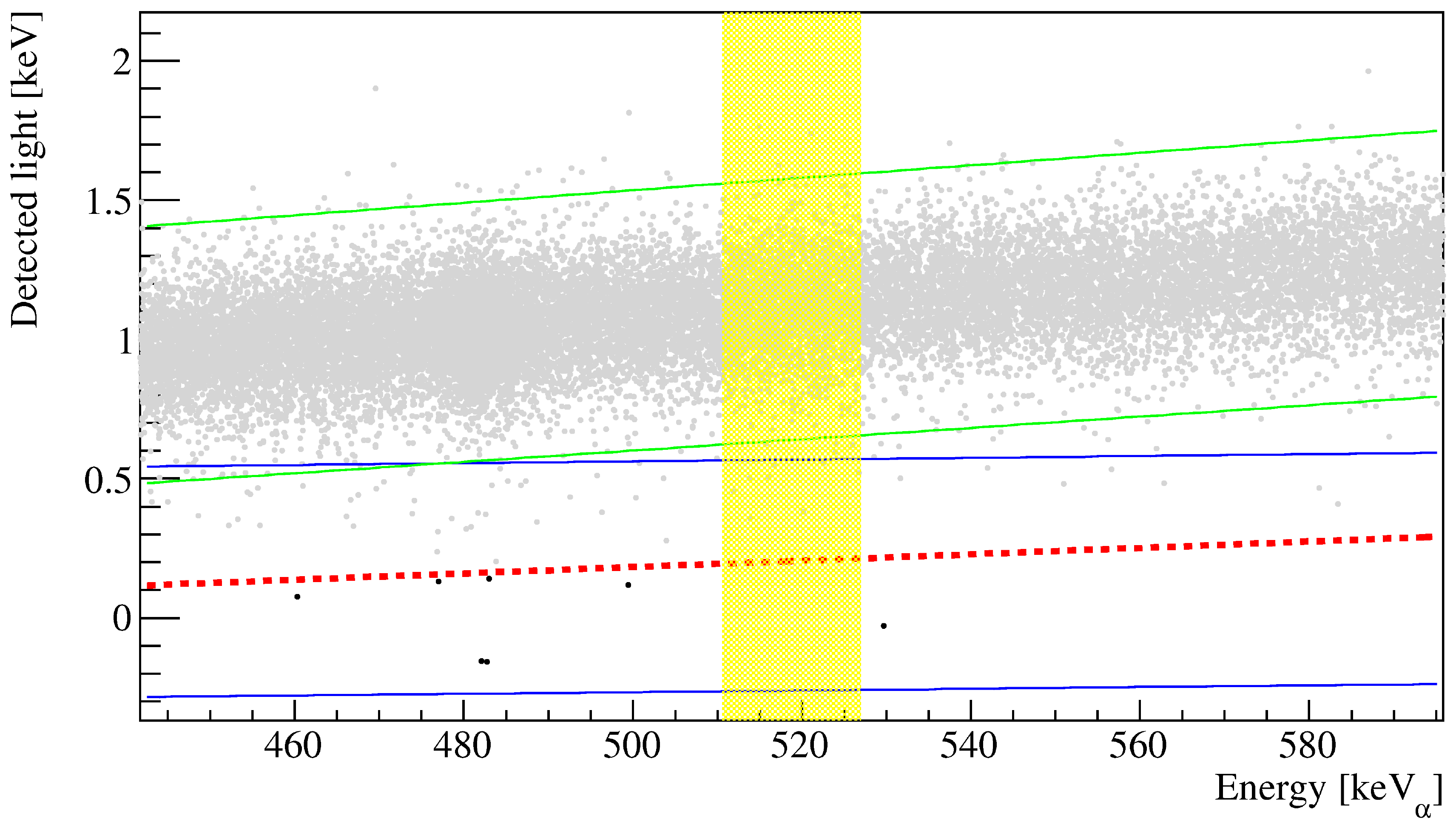}}
\caption{Low energy scatter plot for investigating $^{208}$Pb $\alpha$ decay. The blue and green lines represent the 90\%  acceptance region, respectively, for $\alpha$ and $\beta$/$\gamma$ events. The red dashed line is the energy cut applied for selecting the region with the highest statistical significance, while the shadowed region is our ROI.} 
\label{fig:cut500}   
\end{figure}

The last natural occuring lead isotope to be analyzed is $^{208}$Pb. It can $\alpha$ decay to $^{204}$Hg with an energy transition of 516.9~keV. Also for this nuclide we apply the same procedure previously explained. The studied $\alpha$ band is shown in Fig.~\ref{fig:cut500}, the cut applied on the detected light has an efficiency of 64\%.

The interval centered around the Q-value of the transition is wide 2.8~$\sigma$ (16.4~keV). No events are observed in the ROI, and the estimated background is 0~count. The lower-limit on $^{207}$Pb half-life is:
\begin{equation}
T^\alpha_{1/2}(^{208}Pb)\, >~2.6\cdot10^{21}~y  \; \; \;\mbox{at \;90\% C.L.}
\end{equation}

\section{Conclusion}
We successfully tested a 454.1~g PbWO$_4$ crystal as a bolometer. The crystal was grown using low background ancient Roman lead, and its performances and internal contaminations were presented.\newline

The separation achieved on the scintillation light is a powerful tool that allows us to reject $\beta$/$\gamma$ events for the investigation of low energy rare $\alpha$ decays. The $LY$ of $\alpha$ and $\beta$/$\gamma$ has been calculated.\newline

New more stringent upper limits on the half-lives of four natural lead isotopes has been estimated with a sensitivity of the level of 10$^{20}$-10$^{21}$ y.

\section*{Acknowledgements}
This work was made in the frame of the LUCIFER experiment, funded by the European Research Council under the European Union's Seventh Framework Programme (FP7/2007-2013)/ERC grant agreement no. 247115. Thanks are due to E. Tatananni, A. Rotilio, A. Corsi, B. Romualdi and F. De Amicis for continuous and constructive help in the overall set-up construction.


\begin{thebibliography}{30}

\bibitem{209Bi-nature}
P.~de~Marcillac \emph{et~al.}, Nature \textbf{422}, 876 (2003).

\bibitem{209Bi-nostro}
J.~Beeman \emph{et~al.}, Phys. Rev. Lett. \textbf{108}, 062501 (2012).

\bibitem{cluster-1991}
B.~Buck \emph{et~al.}, J. Phys. G \textbf{17}, 1223 (1991).

\bibitem{cluster-1992}
B.~Buck \emph{et~al.}, J. Phys G \textbf{18}, 143 (1992).

\bibitem{fission-1983}
D.N.~Poenaru, M.~Ivascu, J. Physique \textbf{44}, 791 (1984).

\bibitem{WKB-1992}
B.A.~Brown, Phys. Rev. C \textbf{46}, 811 (1992).

\bibitem{micro-1988}
B.~Al-Bataina, J.~Jaenecke, Phys. Rev. C \textbf{37}, 1667 (1988).

\bibitem{geo}
A.P.~Dickin, \emph{Radiogenic Isotope Geology}, 2nd ed. (Cambridge University Press, Cambridge, 2005).

\bibitem{Riezler-1958}
W.~Riezler, G.~Kauwn, Zeitschrift fur Naturforschung A \textbf{13}, 904 (1958).

\bibitem{Faraggi-1951}
H.~Faraggi, Ann. Phys. \textbf{6}, 325 (1951).

\bibitem{toimass}
http://ie.lbl.gov/toimass.html.

\bibitem{ionizing-1961}
R.D.~Macfarlane, T.P.~Kohman, Phys. Rev \textbf{121}, 1758 (1961).

\bibitem{scint-2003}
F.A.~Danevich \emph{at al}., Nucl. Instr. Meth. A \textbf{556} 259 (2006).

\bibitem{liquid-2009}
K.~Kossert \emph{et~al.}, App. Rad. and Isotopes \textbf{67}, 1702 (2009).

\bibitem{180W-Cresst}
C.~Cozzini \emph{et~al.}, Phys. Rev. C \textbf{70}, 064606 (2004).

\bibitem{AUDI-2003}
G.~Audi, A.H.~Wapstra, C.~Thibault, Nucl. Phys. A \textbf{729}, 337 (2003).

\bibitem{bohlke-2005}
M.~Berglund \emph{et~al.}, Pure Appl. Chem., \textbf{83}, 397 (2011).


\bibitem{lead210-activity-1}
M.~Laubenstein \emph{et~al.}, Appl. Rad. and Isotopes \textbf{61}, 167 (2004).

\bibitem{lead210-activity-2}
K.~Bunzl, W.~Kracke, Nucl. Instr. and Meth. A \textbf{238}, 191 (1985).

\bibitem{Piombo-Romano}
A.~Alessandrello \emph{et~al.}, Nucl. Instr. and Meth. B \textbf{142}, 163 (1998).

\bibitem{PbWO4}
A.A.~Annenkov, M.V.~Korzhik, P.~Lecoq, Nucl. Instr. and Meth. A \textbf{490}, 30(2002).

\bibitem{Birks}
J.B.~Birks, Proc. Phys. Soc. A \textbf{64}, 874 (1951).


\bibitem{Tretyak}
V.I.~Tretyak, Astropart. Phys. \textbf{33}, 40 (2010).


\bibitem{cdwo4}
 C.~Arnaboldi \emph{et~al.}, Astropart. Phys. \textbf{34}, 143 (2010).

\bibitem{lecoq}
P.~Lecoq \emph{et~al.}, Inorganic Scintillators for Detector Systems: Physical Principles and Crystal Engineering, 1st edn. (Springer, Berlin, 2006).

\bibitem{coron}
N.~Coron \emph{et~al.}, Phys. Lett. B \textbf{659}, 113 (2008).

\bibitem {NIMA-2006-A} 
S.~Pirro, C.~Arnaboldi, J.W.~Beeman, G.~Pessina, Nucl. Instr. Meth. A \textbf{559}, 361 (2006).

\bibitem{QINO}
 E.~Andreotti \emph{et~al.}, Astropart. Phys. \textbf{34}, 822 (2011).

\bibitem{NIMA-2006-B} 
S.~Pirro \emph{et~al.}, Nucl. Instr. Meth. A \textbf{444}, 331 (2000).

\bibitem{NIMA-2006-C} 
C.~Arnaboldi, G.~Pessina, S.~Pirro, Nucl. Instr. Meth. A  \textbf{559}, 826 (2006).


\bibitem{GattiManfredi} 
E.~Gatti, P.F.~Manfredi, Rivista del Nuovo Cimento \textbf{9}, 1 (1986).

\bibitem{Radeka:1966} 
V.~Radeka, N.~Karlovac, Nucl. Instr. Meth. A \textbf{52}, 86 (1967).

\bibitem{light_sync}
G.~Piperno, S.~Pirro, V.~Vignati, J. Inst. \textbf{6}, P10005 (2011).
  
\bibitem{CCVR} 
 F.~Alessandria \emph{et~al.}, Astropart. Phys. \textbf{35}, 839-849 (2012).

\bibitem{quencing_factor}
V.I.~Tretyak, Astropart. Phys. \textbf{33} 40-53 (2010).

\bibitem{sticking}
M.~Clemenza, C.~Maiano, L.~Pattavina, E.~Previtali, Eur. Phys. J. C \textbf{71}, 1805 (2011).

\bibitem{Feldman}
G.~Feldman, R.~Cousins, Phys. Rev. D \textbf{57}, 3873 (1998).



\end{thebibliography}
\end{document}